# Nonlinear optical refractive index measurements of pure water via Z-scan technique at 800 nm

**ILYES BETKA**[1*]

[1]*Laboratoire IMS, UMR CNRS 5218, Université de Bordeaux, 33400 Talence, France*
*ilyes.betka@u-bordeaux.fr*

**Abstract:** The determination of the nonlinear optical coefficient $n_2$ of pure water is of great interests for applications such as imaging biological systems, performing femtosecond laser surgery and underwater ablation. Traditionally, probing third-order nonlinearity while eliminating the thermo-optic effect requires high pulse energy and low repetition rates laser sources. In this article, a low cost, simple and versatile third order nonlinear characterization technique for the Z-scan closed aperture mode has been developed to determine the nonlinear refractive index of pure water at 800 nm, with a pump duration of 140 fs and a repetition rate of 80 MHz. Experimental results reveal an $n_2$ value of $(8 \pm 1.4) \times 10^{-20}$ m²/W. The influence of higher-order nonlinearities such as $\chi^5$ was assessed, showing no significant contribution within the examined intensity range. This work paves the way for the use of low power and high repetition rate systems for characterizing third order nonlinear optical materials.

## 1. Introduction

Nonlinear optics is the branch of Optics which describes the interaction of an intense laser with matter [1]. At high intensity levels, the polarization responds non-linearly to the electric field [2] which leads to different phenomena such as Frequency-mixing-processes [3] and other nonlinear process including optical Kerr effect [4] electro-optic effect [5] cross phase modulation [6]. Since the advent of lasers, the study of nonlinear optical materials has become of great interest [7], liquids for example have major applications in imaging biological systems [8], performing femtosecond laser surgery [9] and underwater ablation [10], while Solid state materials have gained more attention because of their use in photonics and the opto-electronic devices [11], especially in optical switching [12], optical communication [13] and optical limiting devices [14].

A key parameter governing most of these applications is the third-order nonlinear susceptibility ($\chi^3$). The latter gives rise to different phenomena like Third harmonic generation [15], optical Kerr effect [4] and optical phase modulation [16]. The real part of this parameter is in return directly related to the nonlinear refraction (NLR) of the material [1]. Many studies and measurement techniques have been devoted to determine this parameter [8] including: Z-scan [17], DFWM [18], beam-distortion measurements [19] and ellipse rotation [20]. Z-scan technique is one of the simplest, accurate [21] and rapid technique to determine the nonlinear absorption coefficient and NLR of different materials [22]. This technique was initially introduced by Sheik-Bahae et al. in 1989-1990 and has since become widely adopted for determining the NLR of nonlinear optical materials [23]. Accurately measuring the $n_2$ value of pure water is crucial for applications such as underwater ablation [10] and femtosecond laser surgery [9]. While several studies have explored the $n_2$ value of pure water around 800 nm, they often face challenges, including reliance on high pump energies up to 100 mJ [24], or the use of alternative techniques to the conventional Z-scan method such as single-shot supercontinuum spectral interferometry (SSSI), which still requires high-energy per pulse excitation [25]. Moreover, in the Z-scan technique, the use of a low repetition rate laser sources is often necessary to eliminate the influence of the thermo-optic effect, which adds complexity to conventional high repetition rate characterization setups. These challenges highlight the need for improved methods that reduce energy consumption and system complexity while ensuring accuracy and reliability. In this work, we introduce a cost-effective, simple, and versatile approach for the Z-scan closed-aperture (CA) method, utilizing a high-repetition-rate, low-energy pump source (several nJ per pulse) at 800 nm. This approach not only overcomes the limitations of previous methods but also provides a more accessible and energy-efficient solution for characterizing third-order nonlinear optical materials.

## 2. Theory of the Z-scan closed aperture technique

Fig 1 shows the working principle of the Z-scan closed aperture technique. This configuration is used to measure the change in the transmittance of the incident Gaussian beam pulse as the sample (thin sample with a length smaller than the diffraction length of the focused beam) is translated through the beam waist [22]. When a Gaussian laser beam profile is focused with a lens, it leads to high intensity at the focal point of the lens and lower intensity elsewhere. An aperture in front of the detector is added in such a way that only the center of the light cone passes through the aperture while the diffraction in the edges of the beam are blocked by the aperture. As the sample is translated through the focused Gaussian beam, it experiences different intensities along the translation path. The intensity dependent nonlinear refractive index arises from the third order nonlinearity of the medium [4]. This latter causes a phase shift which itself manifests a narrowing or a broadening of the beam at the far field (self-focusing/defocusing) [21]. This technique is based on the change of the transmission caused by the phase distortion through the aperture. For a sample with a positive nonlinear refraction coefficient $n_2$, self-focusing effect (Δn > 0) takes place when performing the Z-scan CA technique. This results in a valley followed with a peak in the transmitted field. As the sample moves closer to the beam waist, the self-focusing effect causes the beam to diverge in the far field, leading to a reduction in transmission through the aperture. In the other case, when the sample is positioned after the focus, self-focusing effect causes the beam to converge instead of diverging, this manifests an increase in the transmission through the aperture. A reverse behaviour in the transmission spectrum denotes a negative NLR (Δn <0 ), which is caused by the self-defocusing effect of the nonlinear medium. Therefore, the sign of the NLR is easily readable from the Z-scan experiment. It can be noted that Z-scan CA results in a symmetric curve around Z=0, however although this technique measure only n$_2$, it can be sensitive to measuring both n$_2$ and nonlinear optical absorption coefficient (β). In such cases, the absorption of the medium diminishes the peak and enhances the valley, leading to a slight asymmetry in the curve.

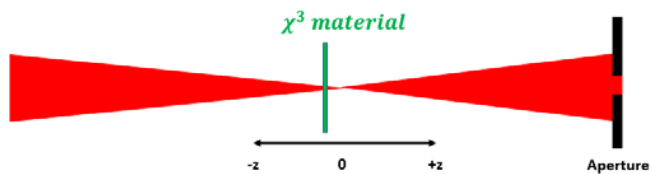
Fig. 1. Z-scan closed aperture configuration

The theory of the Z-scan technique lies behind the induced polarization which depends linearly on the amplitude of the electric field. Thus, when a material is excited with a light beam described by a monochromatic plane wave at the frequency ω, the polarization is defined as :

$$P(\omega) = \varepsilon_0 \chi^{(1)} E(\omega) \quad (1)$$

Where $\chi^{(1)}$ is the first order electrical susceptibility associated with the material considered, $\varepsilon_0$ is the free space permittivity, $E(\omega)$ represents the electric filed. The index of refraction of this material is simply linked to this susceptibility by the relation:

$$n_0^2 = 1 + \chi^{(1)} \quad (2)$$

In order to account for a nonlinear polarization within the material, Taylor series are introduced as follows:

$$P(\omega) = \varepsilon_0 (\chi^{(1)} E(\omega) + \chi^{(2)} E(\omega) E(\omega) + \chi^{(3)} (E(\omega) E(\omega) E(\omega) \dots) \quad (3)$$

$\chi^{(2)}$ and $\chi^{(3)}$ are respectively the second and third order nonlinear electrical susceptibility tensors. The second order nonlinear susceptibility vanishes for media displaying inversion symmetry or so-called centrosymmetric materials. In the other hand, third order nonlinear optical process, described by $\chi^{(3)}$, can occur in both centrosymmetric and non- centrosymmetric media such as amorphous solids, liquids and gasses. Since this work focuses on using pure water as the medium, the nonlinear polarization at the frequency ω for a liquid medium is consequently reduced to:

$$P(\omega) = \varepsilon_0 \chi^{(1)} E(\omega) + 3 \varepsilon_0 \chi^{(3)} |E(\omega)|^2 E(\omega) \quad (4)$$

It is possible to re-formulate this expression in a similar form to that of the linear polarization using an effective susceptibility $\chi_{eff}$ as :

$$\chi_{eff} = \chi^{(1)} + \chi^{(3)} |E(\omega)|^2 \quad (5)$$

The real part of the third order nonlinear susceptibility $\chi_r^{(3)}$ is related to the refraction coefficient called the nonlinear optical refraction $n_2$ :

$$\chi_r^{(3)} = \frac{4 n_0^2 \varepsilon_0 c}{3} n_2 \quad (6)$$

Where c represents the speed of light. The final equation of the refraction index is given by:

$$n(I) = n_0 + n_2 I \quad (7)$$

Where I represent the irradiance of the incident field. The determination of the n₂ value is achieved via the Z-scan closed aperture configuration. The normalized transmittance of the Z-scan CA is expressed as [22] :

$$T(z, \Delta\Phi_0) = 1 - \frac{4 \Delta\Phi_0 \frac{z}{z_0}}{\left[\left(\frac{z}{z_0}\right)^2 + 9\right] + \left[\left(\frac{z}{z_0}\right)^2 + 1\right]} \quad (8)$$

Where z is the position of the position with respect to the focus, $z_0$ represents the Rayleigh rage. $\Delta\Phi_0$ is the phase change at the focus given by :

$$\Delta\Phi_0 = \frac{2\pi}{\lambda} n_2 I L_{eff} \quad (9)$$

$L_{eff}$ represents the sample's effective length. The NLR index can be easily extracted from equation 9.
Typical $n_2$ values of pure water under different excitation wavelength up to 1400 nm are summarized in Fig. 2. This latter has a typical value of several $10^{-20}$ order of magnitude. At wavelength longer than 1150 nm, the reported values start to manifest a high standard deviation value, due to the low beam quality of the pumping source [8]. At approximately 800 nm, two commonly reported $n_2$ values are around $2\times10^{-20}$ and $6\times10^{-20}$ m²/W. These values were obtained using Z-scan [24] and single-shot supercontinuum spectral interferometry, respectively [25]. The latter technique focuses on utilising ultrafast supercontinuum pulses and spectral interferometry to extract nonlinear phase shifts. While this approach enables single-shot measurements and avoids sample translation, it also relies on high-energy per-pulse excitation.

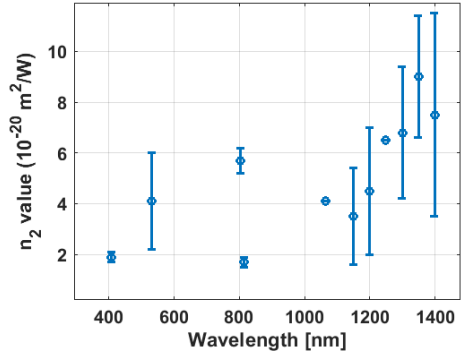

Fig. 2. NLR index of water for different excitation wavelengths from literature

### 3. Experimental setup

The experimental setup is showed in Fig 3. The system consists of a converging lens with a focal length of 12 cm through which a 140 fs, 800 nm with 80 MHz Ti-Sapphire femtosecond laser beam passes. Power and polarization control is achieved via a $\lambda/2$ plate coupled to a cubic polarizer, allowing thus to select the desired power and to fix the state of the incident polarization. The calculated beam waist is 25 µm, and the Rayleigh range was measured to be $z_0 = 2.5$ mm. The liquid water was held by a 5 mm solid container made of silica glass (cuvette). A computer controlled translational stage was used to hold the sample to ensure accurate and automated movement. A diaphragm is placed on the optical path in front of the photodetector to control the aperture size and highlight nonlinear refractive phenomena. For the open aperture configuration, the aperture is fully open (S = 1), where S = 1 represents a completely open aperture. In the closed aperture configuration, the diaphragm is adjusted to S = 0.4 to emphasize the nonlinear refractive effects by restricting the aperture size. The photodiode located at the end of the propagation path measures the intensity transmitted by the sample. An optical density placed just before the photodiode makes it possible to ensure that there is no damage to the photodiode in the event of high power. The sample was irradiated with an irradiance of $I_0 = 4.4$ GW/cm² and the aperture was set to transmit a small portion of the beam (S = 0.4). The transmittance was normalized by dividing each detected value by the first one located on either side of the focal point.

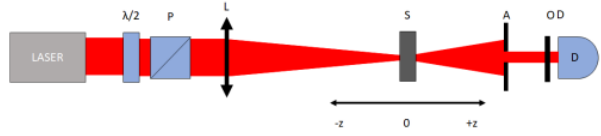

Fig. 3. Experimental setup of the closed aperture Z-scan technique (p: polarizer; $\frac{\lambda}{2}$: half wave plate; L: lens; S: sample; A: aperture; OD: optical density; D: photodiode)

## 4. Results and discussion

Fig. 4 illustrates the obtained results. The black curve represents the experimental data. This latter appears to be relatively symmetric around Z = 0. The graph reveals a valley followed by a peak, suggesting that the sample exhibits a positive nonlinear refractive index (NLR). R-language software was used to fit the data (red curve in Fig. 4) and extract the nonlinear refractive index using equation 9.

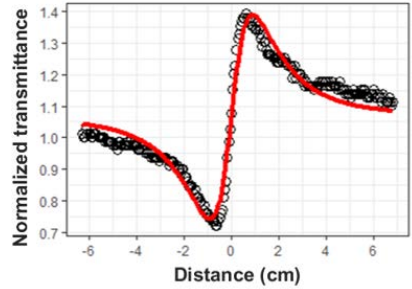

Fig. 4. Z-scan CA results on pure water

The value of $n_2$ was reported to be $n_2$ = (8 ± 1.4) x$10^{-20}$ m$^2$/W. It can be seen that this value is consistent with the values reported in Fig. 2 at 800 nm. Measurements were also conducted on an empty cuvette, yielding a curve similar to the noise background. This indicates that the cuvette itself has no significant impact on the obtained results. It should be noted that often, higher order nonlinearities can take place ($\chi^5$ for example), particularly at higher irradiance levels. To investigate this possibility, measurements were carried out across different irradiance levels [2.2 - 3.08 - 4.4 - 11 GW/cm$^2$], each performed with slightly different values of the beam parameter S. The aim was to assess whether the variation in irradiance would lead to any significant changes in the measured $n_2$ values, potentially revealing the presence of higher-order nonlinearities. The results, shown in Fig. 5, reveal that the variation in $n_2$ values across these measurements is minimal. The consistency of the measured values across the different irradiance levels suggests that the third-order nonlinearity remains dominant within the investigated intensity range, and higher-order nonlinearities, such as $\chi^5$, do not have a noticeable effect under the experimental conditions. This lack of significant variation across a broad range of irradiances supports the conclusion that the contributions from higher-order nonlinearities are negligible for the irradiance levels used in this study.

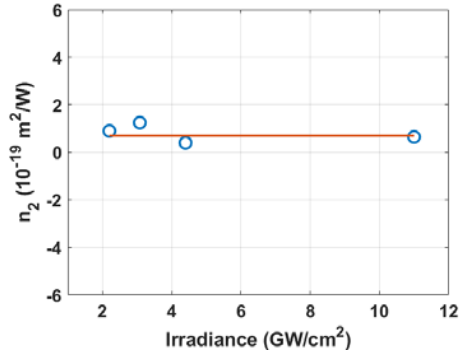

Fig. 5. Measured n₂ values as a function of the irradiance (blue dots represent the experimental data; red curve shows the linear fit )

A comparison of the pumping energy used in this study at 800 nm with previous works conducted at the same wavelength is presented in Fig. 6. A significant difference is observed in the required pumping energy: the experimental results of this work are achieved with a pump energy of 12 nJ, which is orders of magnitude lower than typical values reported in previous studies, such as 1 mJ or 100 mJ.

The key advantage of this approach lies in the combination of low pumping power with high repetition rate. While high repetition rates can introduce the risk of thermo-optic effects, this approach mitigates this concern by using low pump energy. The thermo-optic effect, which depends on both the repetition rate and the intensity, is effectively minimized because the lower pump energy prevents excessive heating of the sample. As a result, the use of low pumping power ensures accurate measurement of the NLR without the complications associated with high energy pulses or the exclusion of thermo-optic effects.

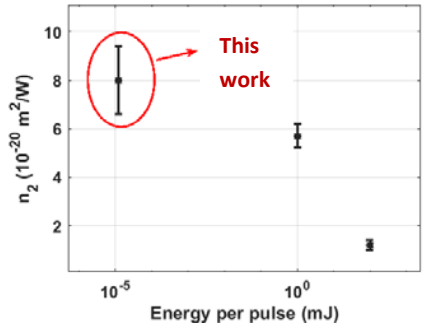

Fig. 6. n₂ values of different studies at 800 nm as a function of the pulse energy

## 5. Conclusion

In conclusion, we demonstrate the effectiveness of the developed Z-scan closed-aperture technique for determining the $n_2$ value of pure water using significantly low pump energy (12 nJ per pulse). The obtained $n_2$ value shows good agreement with values reported in previous studies. Additionally, the potential influence of higher-order nonlinearities was investigated, confirming that they do not significantly affect the results. This work highlights the potential of a simple, precise, and direct method to determine the second-order nonlinear coefficient of nonlinear optical materials using high repetition rate and low-energy pumping lasers. A promising avenue for future work is the integration of a pulse picker [26], which would enable measurements at lower repetition rates for new materials. This would allow for a direct comparison of $n_2$ values obtained at high and low repetition rates, ensuring that the thermo-optic effect does not influence the results. Moreover, the Z-scan experimental setup can be further improved. For example, replacing the photodiode with a camera could provide complete visualization of the deformations in the transmitted laser beam at a given position, thus eliminating the need for a diaphragm. Additionally, implementing a transmission/reflection geometry would enable the investigation of both transparent and highly absorbing nonlinear media. Finally, incorporating broad spectral tuning would allow for a comprehensive spectral characterization, facilitating a more systematic study of various materials.

**Acknowledgment.** The author appreciates Professor Pierre-Francois Brevet, Dr Fabien Rondepierre for their helpful remarks and scientific discussions, and the institute of light and matter for funding.

**Data availability.** Data underlying the results presented in this paper are not publicly available at this time but may be obtained from the authors upon reasonable request.

**Disclosures**. The authors declare no conflicts of interest.